\documentclass[useAMS,usenatbib]{mn2e}

\usepackage{graphics}
\usepackage{rotating}
\usepackage{multirow}

\newcommand{\eg}{{\em e.g.}}
\newcommand{\ie}{{\em i.e.}}

\title[Orbital Elements of Comet C/1490 Y1 and the Quadrantid shower]
{Orbital Elements of Comet C/1490 Y1 and the
Quadrantid Shower}

\author[Lee et al]{Ki-Won~Lee$^{1}$\thanks{email:
kwlee@kasi.re.kr}, Hong-Jin~Yang$^{1}$, Myeong-Gu~Park$^{2}$\\
$^{1}$Korea Astronomy and Space Science Institute, Daejeon 305-348, Korea\\
$^{2}$Department of Astronomy and Atmospheric Sciences, Kyungpook National
University, Daegu 702-701, Korea}
\begin{document}

\date{}

\pagerange{\pageref{firstpage}--\pageref{lastpage}} \pubyear{2009}

\maketitle

\label{firstpage}

\begin{abstract}
The Quadrantid shower, one of the most intense showers, has been
observed at the beginning of January each year.
However, the origin of the meteors is still  
unknown. It was Hasegawa (1979) who first suggested comet C/1490 Y1
to be the likely origin of the shower based on the historical records
of East Asia. 
We analyse the records of {\it Jo-Seon-Wang-Jo-Sil-Lok}
(the Annals of the Joseon Dynasty in ancient Korea) and calculate the
preliminary orbital elements of comet C/1490 Y1 using a modified Gauss method.
We find that comet C/1490 Y1 was
a periodic one and its orbital path was very similar to 
that of the Quadrantid meteor stream.
The determined orbital elements are perifocal passage time 
$Tp$=2265652.2983 days (7.8 Jan. 1491 in UT), perifocal distance 
$q$=0.769 AU, eccentricity $e$=0.747, 
semimajor axis $a$=3.04 AU, argument of the perifocus
$\omega$=164$^{\circ}$.03, longitude of ascending node
$\Omega$=283$^{\circ}$.00, and inclination $i$=70$^{\circ}$.22
for the epoch of J2000.0. We, therefore, conclude that our result
verifies the suggestion that the
comet C/1490 Y1 is the origin of the Quandrantid meteor shower,
but was a periodic comet.
We dicuss a possible link between this comet and the asteroid 2003 EH$_1$
as well. 
\end{abstract}


\section{Introduction}\label{intro}

Meteor shower can be spectacular astronomical events, displaying bright
streaks of light in the sky as the Earth passes through a meteor
stream. One ne of the most regular displays comes from the Quadrantid shower that
can be viewed each January with a Zenithal Hourly rate of about 100. The peak in 
activity is quite short, usually less than a day \citep{shelton65} and the declination
of the radiant is around 50$^{\circ}$, making essentially a Northern Hemisphere shower.
The name of the shower originates from Quadrans Muralis constellation,
which is now a defunct constellation but existed when the stream was 
recognized in 1835 by Quetelet \citep{fisher30}. 
Up to present, however, the parent body of the meteors has not been 
clearly identified \citep{kanuchova07}.

There have been numerous suggestions regarding a possible parent for the stream,
starting with \cite{bouska53} who suggested comets C/1939 B1 and 8P/1790 A2 as 
possibilities. A list of early suggestions can be found in \cite{williams04b}.
More recently, comet 99P/ Macholz has been discussed by many authors 
\citep{mcintosh90, jones93, babadzhanov91, kanuchova07}. The increase in the number 
of Near Earth asteroids discovered has also lead to some being identified as having
similar orbits to the Quadrantids, for example 1973 NA \citep{williams98} and 2003 EH$_1$
\citep{jenniskens04}. \cite{babadzhanov08} showed that the Quadrantids and 2003 EH$_1$ 
are definitely related. An other strong candidate, first suggested by 
\cite{hasegawa79}, is comet C/1490 Y1. \cite{williams04a} discussed the possibility that
this comet and asteroid 2003 EH1 are dynamically related. Hasegawa's suggestion is 
based on the historical records from China, Korea and Japan of the appearance of 
the comet. As in most other studies on ancient comets, he assumed the orbit of the 
comet to be parabolic (\ie eccentricity of unity). Except for the eccentricity,
his orbital path showed a good agreement with that of the Quadrantids 
\citep{williams93}. More details reviews of comet C/1490 Y1 and of the Quadrantids
can be found in \cite{kronk99} and \cite{jenniskens08}.

\begin{table*}
\begin{minipage}[!hpt]{16cm}
\begin{center}
\caption[Summary of the records]
{Summary of some records related to comet C/1490 Y1 from Sillok.}
\label{table1}
\begin{tabular}{lccl}
\hline
\hline
\multicolumn{3}{c}{Date} & \multicolumn{1}{c}{Records} \\
\hline
 3 & Jan. & 1491 & One-Watch$^a$ yesterday, there was a faint light
     with 4$\sim$5 {\em chi}$^b$ tails around the 11th lunar mansion. \\
   &     &      & King Seongjong ordered observation to Eung-Gi Kim 
                    and Ji-Seo Jo. \\
 6 & Jan. & 1491 & One-Watch yesterday, the comet was
                   unobservable due to heavy clouds. \\
 7 & Jan. & 1491 & Yesterday, the position of the comet was 6$^{\circ}$ 
    from the 12th lunar mansion and 65$^{\circ}$ from the north pole. \\
   &     &       & Identified as a comet due to its tail and the direction. \\
 8 & Jan. & 1491 & Venus appeared at the daytime.\\
 9 & Jan. & 1491 & Yesterday, the position of the comet was 11$^{\circ}$ 
     from the 12th lunar mansion and 76$^{\circ}$ from the north pole.\\
   &     &      & Used Simplified Astronomical Instrument
                  for the observations of the comet. \\
10 & Jan. & 1491 & Yesterday, the position of the comet was 14$^{\circ}$
                  from the 12th lunar mansion and 79$^{\circ}$
                  from the north pole. \\
11 & Jan. & 1491 & Yesterday, the position of the comet was 2$^{\circ}$
         from the 13th lunar mansion and 81$^{\circ}$ from the north pole. \\
12 & Jan. & 1491 & Yesterday, the position of the comet was 4.5$^{\circ}$
      from the 13th lunar mansion and 84$^{\circ}$ from the north pole. \\
14 & Jan. & 1491 & Yesterday, the comet was located above Yun-Yu$^c$ 
                   constellation. \\
15 & Jan. & 1491 & Yesterday, the comet moved into below the 1st star 
                of Yun-Yu constellation from the east. \\
20 & Jan. & 1491 & Yesterday, the comet moved into between Lei-Pi-Chhen$^c$
                   and Thien-Hun$^c$ constellations. \\
21 & Jan. & 1491 & Yesterday, the comet moved into above the 1st star of
          Thien-Tshang$^c$ constellation from the west ($\iota$ Cet). \\
23 & Jan. & 1491 & Yesterday, the comet trespassed the 2nd star of 
     Thien-Tshang constellation from the west ($\eta$ Cet). \\
25 & Jan. & 1491 & First-Watch$^a$ yesterday, the comet moved into the centre
                of Thien-Tshang constellation. \\
26 & Jan. & 1491 & First-Watch yesterday, the comet moved into 2 or 3 {\em chi}
                south-west of the 2nd star of Thien-Tshang \\ 
   &      &      & constellation from the east ($\tau$ Cet). \\
29 & Jan. & 1491 & First-Watch yesterday, the comet moved into 2 or 3 {\em chi} 
                 south-east of $\tau$ Cet. \\
31 & Jan. & 1491 & One-Watch yesterday, the comet trespassed the 1st star
       of Thien-Tshang constellation from the east (57 Cet). \\
05 & Feb. & 1491 & First-Watch yesterday, the comet moved into the east
                   of 57 Cet. \\
   &     &      & A star appeared from the east of $\eta$ Cet. \\
06 & Feb. & 1491 & One-Watch yesterday, the comet moved into the east of
                   57 Cet.\\
   &     &      & A star moved into the east of 3rd star of Thien-Tshang
                  constellation from the west ($\theta$ Cet). \\
09 & Feb. & 1491 & The Comet disappeared. \\
   &     &       & A star moved into the west of Thien-Tshang 
                   constellation. \\
12 & Feb & 1491 & A star disappeared. \\
\hline
\hline
\multicolumn{4}{l}{$^a$refer to text} \\
\multicolumn{4}{l}{$^b$1 {\em chi} = 1.5$^{\circ} \pm 0.24^{\circ}$
\citep{kiang72}}\\
\multicolumn{4}{l}{$^c$refer to Fig.~\ref{fig1}}
\end{tabular}
\end{center}
\end{minipage}
\end{table*}

For Korean records of 
comet C/1490 Y1, \cite{hasegawa79} referred to
Ho's (1962) catalogue which was compiled
from {\it Jeung-Bo-Mun-Heon-Bi-Go} (Explanary Notes of Literary Document).
This is a secondary historical document
compiled from the various first historical ones 
but abbreviated in descriptions, and therefore contains some errors.
Nonetheless, the book was one of the most frequently cited references on 
the study of ancient Korean astronomy \citep[\eg,][]{rufus36, needham86}
because it deals with various astronomical subjects
covering the whole period of historic Korea.

In this paper, we examine the records on comet C/1490 Y1 from
{\it Jo-Seon-Wang-Jo-Sil-Lok} (the Annals of the Joseon Dynasty 
in Korea [1392 -- 1910]; Sillok, hereafter) and calculate the orbital elements
using a modified Gauss method.
In section~\ref{records}, we list some relevant records
on comet C/1490 Y1 from Sillok and
briefly introduce the ancient astronomy in Korea. In section~\ref{analysis},
we explain the process of determining preliminary
orbital elements from the records of Sillok. We present our
results in section~\ref{result} and
summarise in section \ref{summary}. 

\section{Records of Comet C/1490 Y1 from Sillok}\label{records}
Different from {\it Jeung-Bo-Mun-Heon-Bi-Go},
Sillok is one of the first-hand historical documents and contains 
more detailed descriptions of astronomical phenomena. In some cases, 
Sillok's records contain additional information 
such as observation time,
angular distance from the north pole, length of a tail (in case of
a comet), and so forth. 
Hence, Sillok is one of the most important sources on the studies of
the past astronomical phenomena
\citep[\eg,][]{stephenson87, hasegawa03, yang03}.
Recognized for its importance, 
Sillok was registered as a Memory of the World 
by UNESCO in 1997.

\begin{figure*}
\begin{minipage}[hpt]{17cm}
\begin{center}
\includegraphics[scale=0.9,angle=0,clip=true]{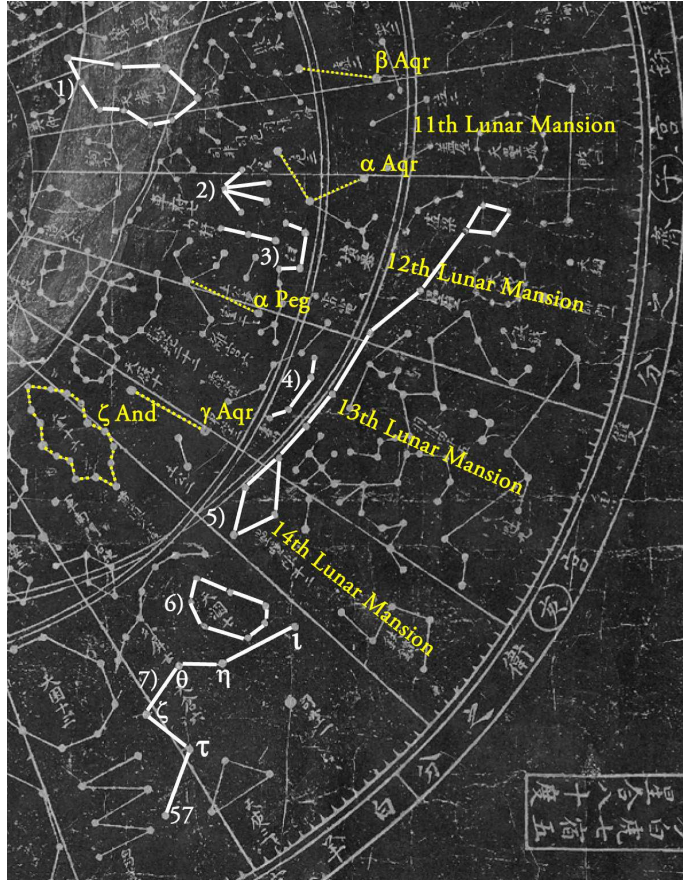}
\caption{\label{fig1} Oriental constellations presented in
{\it Cheon-Sang-Yeol-Cha-Bun-Ya-Ji-Do}, ancient Korean star chart
engraved on a stone.  Constellations connected by
white-solid lines are 1) Thien-Chin, 2) Jen-Hsing,
3) Chhu \& Chiu, 4) Yun-Yu, 5) Lei-Pi-Chen, 6) Thien-Hun,
and 7) Thien-Tshang. While, lunar mansions are depicted by yellow-dotted
lines together with referecne stars in each mansion
(Courtesy of Kyujanggak Library).
}
\end{center}
\end{minipage}
\end{figure*}

Chinese astronomer first discovered comet C/1490 Y1
at the south of Thien-Chin constellation on the end of 1490 and
reported last on January 30 next year \citep{kronk99}.
According to Sillok, this comet had been observed for
39 days nearly every day from 2 Jan. to 9 Feb. 1491 in Korea.
In Table~\ref{table1},
we summarise some relevant records presented in Sillok.
The first and the second columns are 
observation dates (in Gregorian calendar) and the contents of
Sillok's records, respectively. 
The table shows how systematic and detailed the records of Sillok are.
For example, the record of 23 Jan. 1419 tells us 
that comet C/1490 Y1 trespassed against $\eta$ Cet on
22 Jan. 1491
while Chinese record simply says that
`comet trespassed against Thien-Tshang' \citep[see][]{ho62}.
In addition, the Annals also describe
some interesting facts such as who performed the observations at that
time (i.e., Eung-Gi Kim and Ji-Seo Jo), why it was identified as a comet
(i.e., due to the tail and its direction), what kind of an astronomical
instrument was used in observations (i.e., Small Simplified Astronomical
Instrument; see Needham et al. 1986), and how to operate the instrument (not
present in this study). Lastly, 
a star appeared after 5 Feb. 1491 
is object designated as X/1491 B1 in Kronk's (1999) catalogue.
We express the names of the oriental constellation and the units of
the angular distance in terms of the Chinese transliteration
because those are better known to international community.
In Figure~\ref{fig1}, we
present a part of {\it Cheon-Sang-Yeol-Cha-Bun-Ya-Ji-Do}
\citep[an ancient Korean star chart engraved on a stone in 1395, refer to]
[]{rufus12,needham66,park98} in order to provide a visual
impression on oriental constellations and to help the understanding
of ancient Korean uranography explained in the next section.
In the figure, yellow-dotted lines are lunar mansions and
$\beta$ Aqr, $\alpha$ Aqr, $\alpha$ Peg, $\gamma$ Aqr, and $\zeta$ And
are reference stars in each mansion.
While white-solid one are
oriental constellations 
mentioned in this
study, 1) Thien-Chin, 2) Jen-Hsing, 3) Chhu \& Chiu, 4) Yun-Yu,
5) Lei-Pi-Chhen, 6) Thien-Hun, and 7) Thien-Tshang.

Before 1895, the history of calendrical methods in Korea is
basically the same as in China, hence, the history of hour systems
as well. Until 1653,
the Joseon royal court used the 100 Divisions system
which divides the day into one hundred degrees.
The court also enforced Five Watches system for the nighttime;
the nighttime, period
from 2.5 degrees (\ie, 0.6 hours) after sunset
to 2.5 degrees before sunrise the next morning, was equally divided
into five intervals. In this hour system, One Watch was interchangeably 
called First Watch and each Watch also was subdivided into five intervals.
Of necessity, each Watch has different time length every day.
To take account of twilight hours, 2.5 degrees was introduced into
the system of night hours.
More details on calendrical methods and time measuring systems can 
found in \citet{jeon74}. 

\section{Data reduction}\label{analysis}
\subsection{Record analysis}
Like in ancient Chinese astronomy, 
the Joseon astronomers divided the whole sky into 28 lunar mansions
and each lunar mansion starts from a reference star in right ascension.
For example, the 11th lunar mansion begins with $\beta$ Aqr
(see Fig.~\ref{fig1}).
As can be found in Table~\ref{table1}, all observational records
by an astronomical instrument are described in the form of
the angular distances both from a lunar mansion and the north pole.
However, the value of
365.25 degrees was used to cover the great circle
in the sky until 1650s.
Therefore, the observational values by an astronomical instrument 
simply can be converted into right ascension and declination as the followings:
\begin{eqnarray}
\label{eq1}
\alpha &=& \alpha_k + C \times 
\frac{\rm ADL}{15}\\ \nonumber
\delta &=& 90^{\circ} - C \times 
{\rm ADN},
\end{eqnarray}

\begin{table*}
\begin{minipage}[!hpb]{16cm}
\begin{center}
\caption[Angular position for comet C/1490 Y1]
{Angular position data for comet C/1490 Y1 in the epoch of 1491.0}
\label{table2}
\begin{tabular}{ccccccc}
\hline
\hline
ID & Julian Day          & $\alpha_{1491.0}$ & $\delta_{1491.0}$ & 
X & Y & Z \\
\cline{5-7}
   & (UT) & (hours)           & (degrees)         &
\multicolumn{3}{c}{(AU)}\\
\hline
 1 & 2265650.884819 & 22.053676 & 25.93429 & 0.4176779 & $-$0.8172441 &
$-$0.3554515 \\ 
 2 & 2265652.886257 & 22.382218 & 14.59959 & 0.4491863 & $-$0.8033028 &
$-$0.3493867 \\ 
 3 & 2265653.886987 & 22.579343 & 12.13552 & 0.4647339 & $-$0.7959541 &
$-$0.3461899 \\ 
 4 & 2265654.887722 & 22.789650 & 10.16427 & 0.4801369 & $-$0.7883565 &
$-$0.3428849 \\
 5 & 2265655.888462 & 22.953921 &  7.20739 & 0.4953900 & $-$0.7805127 &
$-$0.3394730 \\
 6 & 2265666.896796 &  0.716267 & $-$12.91460 & 0.6519516 & $-$0.6788436 &
$-$0.2952586 \\
 7 & 2265674.902872 &  1.596609 & $-$23.35432 & 0.7508476 & $-$0.5888331 &
$-$0.2561119 \\
\hline
\hline
\end{tabular}
\end{center}
\end{minipage}
\end{table*}
where $\alpha_k$ is the right ascension of the reference star of the
$k$th lunar mansion in the unit of hours and $C$ is the correction term of 
$\frac{360}{365.25}$. 
ADL and ADN represent angular distances from a lunar mansion and
from the north pole, respectively, in the unit of degrees.
In this study, we use $\beta$ Aqr, $\alpha$ Aqr, and  $\alpha$ Peg as
the reference stars of the 11th, 12th, and 13th lunar mansions,
respectively \citep{kiang72, cullen06}.

Besides five records that include observational values,
Sillok also provides two valuable information
on the comet: trespass events on 23 and 31 Jan. 1491.
According to {\em Seo-Un-Gwan-Ji}, Records of {\em Seo-Un-Gwan}
(the Royal Bureau of Astronomy in the Joseon dynasty), a tresspass
event is defining as a phenomenon in which the light ray of two
celestial bodies has influence on each other within a {\em cun},
a tenth of a {\em chi} (\ie, 0.15 degrees).
We, therefore, use the coordinates of invaded stars 
for the records of a trespass event.
To identify the invaded stars,
we refer to the work of \citep{ahn96} 
They studied oriental stars listed in {\it Seong-Kyung},
a star catalogue written by a Korean astronomer Byeong-Gil
Nam in 1861, to identify with modern ones.
According to their work, the 1st
star of Thien-Tshang constellation from the east is 57 Cet not
$\upsilon$ Cet \citep[cf.][]{kiang72}.
For the purpose of illustration,
we also depict reference stars in Fig.~\ref{fig1} together with the stars
addressed in this study.

\subsection{Data reduction}
\label{reduction}
\begin{table*}
\begin{minipage}[!hpt]{16cm}
\begin{center}
\caption[Orbital elements of comet C/1490 Y1]
{Orbital elements of comet C/1490 Y1 and associated objects
in the epoch of J2000.0}
\label{table3}
\begin{tabular}{lccccccc}
\hline
\hline
\multicolumn{1}{c}{Objects} &$T_P$&$q$&$e$& $a$ &$\omega$&$\Omega$&$i$ \\
    & (UT) & (AU) &  & (AU) & (deg.) & (deg.) & (deg) \\
\hline
Quadrantids (1995)$^a$ & $\cdots$ & 0.979 & 0.69 & 3.14 & 171.2 & 283.1 &
71.05+72.7 \\
Quadrantids (1491)$^b$ & $\cdots$ & 0.758 & 0.74 & 2.94 & 164.5 & 284.7 &
70.5 \\
C/1490 Y1 (1491)$^c$ & 7.8 Jan. 1491 & 0.769 & 0.75 & 3.04 & 164.0 & 283.0 & 70.2 \\
C/1490 Y1 (1491)$^d$ & 8.9 Jan. 1491 & 0.761 & 1.00 & $\cdots$ & 164.9 & 280.2 & 73.4\\
2003 EH$_1$ (1491)$^e$ &  13 Apr. 1490  & 0.570 & 0.82 & 3.17 & 164.2 & 
286.2 & 66.0 \\
2003 EH$_1$ (1491)$^a$ & (8.9 Jan. 1491) & 0.759 & 0.76 & 3.10 & 164.5 & 
285.5 & 69.2 \\
2003 EH$_1$ (1491)$^b$ & $\cdots$ & 0.732 & 0.76 & 3.10 & 164.0 & 285.5 &
 67.6 \\
2003 EH$_1$ (2003)$^a$ &  24.5 Feb. 2003  & 1.192 & 0.62 & 3.13 & 171.4 &
282.9 & 70.8 \\
\hline
\hline
\multicolumn{8}{l}{$^a$\cite{jenniskens04},~$^b$\citet{williams04b}
\cite[see also][]{wu92,williams93},
~$^c$this study,} \\
\multicolumn{8}{l}{$^d$\cite{hasegawa79}, $^e$\cite{micheli08}}
\end{tabular}
\end{center}
\end{minipage}
\end{table*}
We use total seven data points in orbital calculations: records on
7, 9, 10, 11, 12, 23 and 31 Jan. 1491. 
We compute
the coordinates of the reference stars
for the epoch of 1491.0 by 
correcting the effects of the precession and nutation
and then calculate the right ascensions and
declinations of comet C/1490 Y1 using Eq.~\ref{eq1}.
We use the algorithms of \cite{meeus98} and
proper motion values of Perryman et al. (1997).
To estimate the observation time, we firstly
subtract one day from the day of the record
because all records are mentioning the observations
performed in the previous day.
Next, we compute the time span ranging from the
sunset to the sunrise using Meeus' algorithms \citep{meeus98}
and VSOP87 solutions with the full periodic terms
\citep[refer to][]{bretagnon88}.
By subtracting 1.2 hours from the time span
and dividing into five intervals, we estimate the time corresponding to
One-Watch. For the records without information on 
the observation time,
we assume One-Watch by analogy with the remaining records.

In Table 2, we summarise seven angular position data for comet 
C/1490 Y1. The first column is an identification number, the second one
the observation time in UT, the third one the right ascension in hours, 
and the forth one the declination in degrees. The last three columns are 
rectangular coordinates of the Sun in AU.

\subsection{Error estimate}
\label{error}
Following the records of comet C/1490 Y1, there are also records
of the observations for the Mars from 5 March to 2 May 1491.
We compared the reported positions with the modern celestrial
mechanical calculations
and found positional errors of
$0.080\pm 0.042$ hours and $0.310 \pm 0.040$ degrees in right
ascension and declination, respectively.
For the records of trespass events,
the error is 0.15 degrees in angular distance as mentioned above.
We also found that each Watch had the range of $\sim$ 2 hours around January.

\section{Results}
\label{result}
For 35 subsets made from the combination of three out of seven data,
we calculate preliminary orbital elements
using the modified Gauss method developed by
\cite{marsden85,marsden91} that uses
coordinate transformation and closed f and g series. 
Of 35 subsets, we obtain the orbital elements similar to
the Quadrantids stream from the
subset composed of records on 11, 23, and 31 Jan. 1491. 
This result indicates that the records
of trespass events (\ie, 23 and 31 Jan. 1491) contribute most to 
the orbital calculation.
It is a natural consequence
because the records of trespass events provide more accurate 
information on comet's 
positions than those of the observations by an astronomical instrument.
Trespass events are also very useful because they
cover the longest time span compared with the first four data in Table~2,
which are very close in time and, therefore, essentially providing
only one point on the orbit. 
Julian day of  2265660 in Table~2 corresponds to mid day
on 15 Jan. 1491.

We performed a least square adjustment
\citep[\eg,][]{boulet91} to improve the orbital elements
with all seven data points. However,
we failed presumably due to
somewhat large errors in observations by an astronomical
instrument, particularly in right ascension
($\sim$ 1.2$^{\circ}$). In Table~\ref{table3},
we present our and Hasegawa's preliminary orbital elements for comet
C/1490 Y1. We also list the results of other studies on associated
objects.  In the Table, the first column is an object name and 
other columns are orbital elements:
$T_P$ is the time of passage in perihelion, $q$ perihelion distance,
$e$ eccentricity, $a$ semimajor axis, $\omega$ argument of
perihelion, $\Omega$ longitude of ascending node, and $i$ inclination.
All elements are reduced to the equinox J2000.0. 
Except for the eccentricity, 
our orbital elements match well with Hasegawa's overall
confirming the correctness of our calculation.
As can be readily seen in the Table, however, 
our result show a better agreement than Hasegawa's
with those of the Quadrantids of the year 1491
derived by \cite{williams04b}.

\begin{figure*}
\begin{minipage}[hpt]{17cm}
\begin{center}
\includegraphics[scale=0.6,angle=0,clip=true]{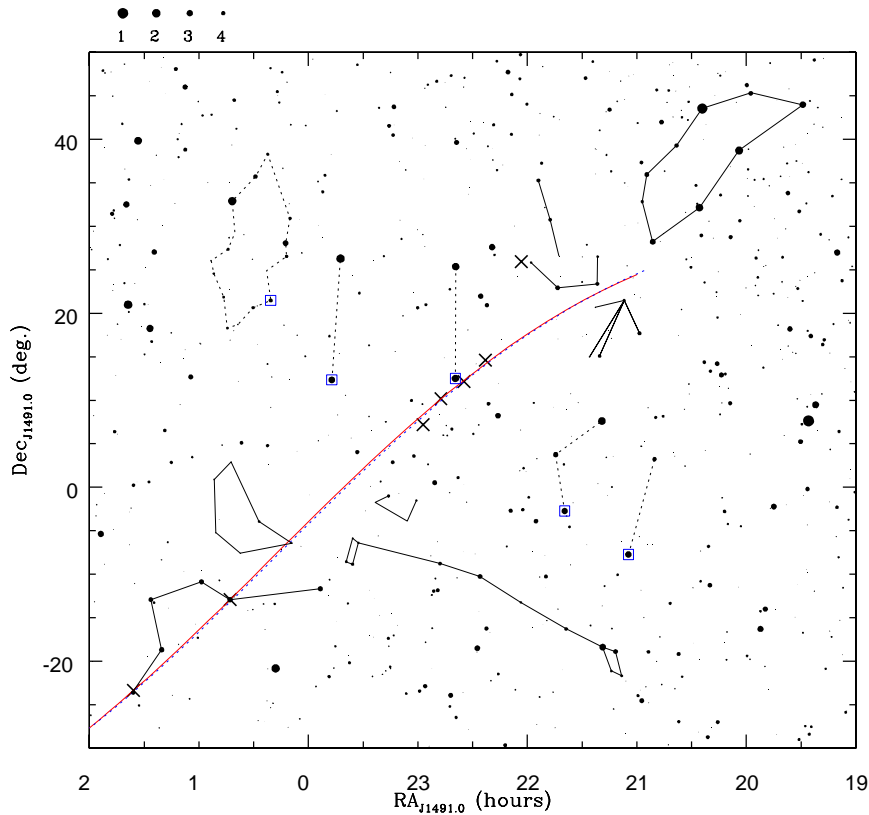}
\caption{\label{fig2} The path of comet C/1490 Y1
together with oriental constellations (solid line) and
lunar mansions (dotted line).
The red-solid and blue-dotted lines are orbital paths from
our and Hasegawa's studies, respectively.
The cross symbols represent seven data points from Sillok's records
and filled blue-rectangulars are reference stars in each lunar mansion.}
\end{center}
\end{minipage}
\end{figure*}

In Fig. \ref{fig2}, we depict the paths of comet C/1490 Y1 using our 
and Hasegawa's orbital elements. The cross symbols 
represent comet's positions from Sillok's records.
The red-solid and blue-dotted
lines show comet's paths from our and Hasegawa's orbits.
Because the comet was observed around the perihelion,
it is hard to distinguish between both paths despite the
different eccentricities.  Although both paths closely pass through
Jen-Hsing, they can not satisfy the Chinese record, which described
the constellation the comet trespassed \citep{ho62}.
In Fig~\ref{fig2}, we also present oriental constellations, lunar mansions,
and reference stars for the purpose of the comparison with
Fig. \ref{fig1}. 

The most remarkable point in our result is the eccentricity.
Our orbital elements
show that comet C/1490 Y1 is a periodic one. However, it
has been never observed since 1491. Assuming a nodal distance of 5.2 AU
in Hasegawa's orbital elements (hence, $e$=0.768 and $a$=3.28 AU),
\citet{williams93} suggested 
a possibility that the comet escaped from 
the observable orbit or was broken by the perturbation of the Jupiter. 
However, our result (\ie, nodal distance of 7.1 AU) suggests
no strong interaction with the Sun and/or planets.
Therefore, the result support the suggestion of
\cite{jenniskens04} in which the parent body of the Quadrantids would
still remain.
One possible hypothesis is that the comet
was originally an asteroid but showed the cometary
activity around 1491 but now the activity turned off
and remains as a dormant comet.
Though we can not explain what caused the activity,
several minor planets which displayed cometary activity
are currently reported \citep{groussin04, hsieh09}.
The orbital elements, of course, would change as they lose
a mass.
This hypothesis are also good accords with the suggestion of 
\cite{jenniskens04}:
the Quadrantid stream originated from comet
C/1490 Y1 around 500 years ago and presumably asteroid 2003 EH$_1$
is the comet's remnant.

Meanwhile, Sillok also contains doubtful records relating to
comet C/1490 Y1. If our (or Hasegawa's) orbital path is correct,
the angular distance from the north pole is $\sim$ 74 degrees (in Chinese
degree) not 65 as recorded on 7 Jan. 1491; the observational value
might be a typo.
If the record is true, on the other hand, it means that the comet
underwent a violent change when approaching to the perihelion.
As an another example, there is
no record of comet C/1490 Y1 on the 8th January although the daytime
appearance of the Venus is recorded. 
In this case, it seems that the record was 
merely omitted rather than the comet was actually unseen
because Sillok does not clearly state that the comet was unobservable
as in the record of 6 Jan. 1491.
However, we can not exclude the possibility that
the comet was actually unobservable because it passed through
the perihelion at that time. In other words, the possibility that 
the true perihelion
passage time is about 7.3 days in UT (remember all records
are writing on the events that had happened the day before).

\section{Summary}\label{summary}
We investigate the astronomical records of Sillok and
calculate the orbital elements of 
comet C/1490 Y1.  Sillok contains valuable information on the comet such as
observation time, right ascension, declination, and so on.
However, it has never been directly referred in previous studies.
Using the modified Gauss method, we compute preliminary
orbital elements of the comet based on the observational records in Sillok.
We find out that our
orbital elements show an excellent agreement with 
those of the Quadrantid stream including the eccentricity.
This fact strongly supports that
comet C/1490 Y1 is the origin of the Quadrantid shower as
firstly pointed out by \cite{hasegawa79}.
Also, our result shows that the comet was a periodic one and
had no experiences of closer encounter with the Sun or planets.
Nonetheless, it is known that the comet
has never been observed before and after 1491. So we suspect that
the comet still remains as a dormant comet such as 2003 EH$_1$ on a
different orbital path from that of
the parent body due to the cometary activity.

\section*{Acknowledgments}
We thank the referee, Iwan Williams, for his kind
and helpful comments.
KWL and MGP are partially supported by Korea Science and Engineering
Foundation through the Astrophysical Research Center for the Structure
and Evolution of the Cosmos (ARCSEC) of Sejong University. HJP and
MGP are supported by Korea Astronomy and Space Science Institute.

\bibliographystyle{mn2e} 

\bibliography{c1491} 

\begin{thebibliography}{}

\bibitem[\protect\citeauthoryear{Ahn, Park \& Yu}{Ahn et~al.}{1996}]{ahn96}
Ahn S.,  Park C.,    Yu K.~L.,  1996, Journal of the Korean History of Science
  Society, 18, 3

\bibitem[\protect\citeauthoryear{Babadzhanov \& Obrubov}{Babadzhanov \&
  Obrubov}{1991}]{babadzhanov91}
Babadzhanov P.~B.,  Obrubov {\rm Yu}.~V.,  1991, in Harris, A. W. and Bowell,
  E. (ed.), Asteriods, Comets, Meteors.
Tucson: Lunar and Planetary Inst.

\bibitem[\protect\citeauthoryear{Babadzhanov, Williams \&
  Kokhirova}{Babadzhanov et~al.}{2008}]{babadzhanov08}
Babadzhanov P.~B.,  Williams I.~P.,    Kokhirova G.~I.,  2008, \mnras, 386,
  2271

\bibitem[\protect\citeauthoryear{Boulet}{Boulet}{1991}]{boulet91}
Boulet D.~A.,  1991, Methods of Orbit Determination for the Microcomputer.
Willmann-Bell Inc.

\bibitem[\protect\citeauthoryear{Bou\v{s}ka}{Bou\v{s}ka}{1953}]{bouska53}
Bou\v{s}ka J.,  1953, Bull. Astron. Inst. Czechosl., 4, 165

\bibitem[\protect\citeauthoryear{Bretagnon \& Francou}{Bretagnon \&
  Francou}{1988}]{bretagnon88}
Bretagnon P.,  Francou G.,  1988, \ana, 202, 309

\bibitem[\protect\citeauthoryear{Cullen}{Cullen}{2006}]{cullen06}
Cullen C.,  2006, Astronomy and mathematics in ancient China: the {\em Zhou bi
  suan jing}.
Cambridge Univ. Press

\bibitem[\protect\citeauthoryear{Fisher}{Fisher}{1930}]{fisher30}
Fisher I.~W.,  1930, Harvard Coll. Obs. Circ., No. 346

\bibitem[\protect\citeauthoryear{Groussin, Lamy \& Jorda}{Groussin
  et~al.}{2004}]{groussin04}
Groussin O.,  Lamy P.,    Jorda L.,  2004, \aa, 413, 1163

\bibitem[\protect\citeauthoryear{Hasegawa}{Hasegawa}{1979}]{hasegawa79}
Hasegawa I.,  1979, \pasj, 31, 257

\bibitem[\protect\citeauthoryear{Hasegawa \& Nakano}{Hasegawa \&
  Nakano}{2003}]{hasegawa03}
Hasegawa I.,  Nakano S.,  2003, \mnras, 345, 883

\bibitem[\protect\citeauthoryear{Ho}{Ho}{1962}]{ho62}
Ho P.~Y.,  1962, \va, 5, 127

\bibitem[\protect\citeauthoryear{Hsieh, Jewitt \& Ishiguro}{Hsieh
  et~al.}{2009}]{hsieh09}
Hsieh H.~H.,  Jewitt D.,    Ishiguro M.,  2009, \aj, 137, 157

\bibitem[\protect\citeauthoryear{Jenniskens}{Jenniskens}{2004}]{jenniskens04}
Jenniskens P.,  2004, \aj, 127, 3018

\bibitem[\protect\citeauthoryear{Jenniskens}{Jenniskens}{2008}]{jenniskens08}
Jenniskens P.,  2008, Meteor Showers and their Parent Comets.
Cambridge Univ. Press

\bibitem[\protect\citeauthoryear{Jeon}{Jeon}{1974}]{jeon74}
Jeon S.-W.,  1974, Science and Technology in Korea: Traditional Instruments and
  Techniques.
MIT Press

\bibitem[\protect\citeauthoryear{Jones \& Jones}{Jones \&
  Jones}{1993}]{jones93}
Jones J.,  Jones W.,  1993, \mnras, 261, 605

\bibitem[\protect\citeauthoryear{Ka\v{n}uchov\'{a} \&
  Neslu\v{s}an}{Ka\v{n}uchov\'{a} \& Neslu\v{s}an}{2007}]{kanuchova07}
Ka\v{n}uchov\'{a} Z.,  Neslu\v{s}an L.,  2007, \ana, 470, 1123

\bibitem[\protect\citeauthoryear{Kiang}{Kiang}{1972}]{kiang72}
Kiang T.,  1972, Mem. R. Astr. Soc., 76, 27

\bibitem[\protect\citeauthoryear{Kronk}{Kronk}{1999}]{kronk99}
Kronk G.~W.,  1999, Cometography: A Catalog of Comets.
Cambridge Univ. Press.

\bibitem[\protect\citeauthoryear{McIntosh}{McIntosh}{1990}]{mcintosh90}
McIntosh B.~A.,  1990, Icarus, 86, 299

\bibitem[\protect\citeauthoryear{Marsden}{Marsden}{1985}]{marsden85}
Marsden B.~G.,  1985, \aj, 90, 1541

\bibitem[\protect\citeauthoryear{Marsden}{Marsden}{1991}]{marsden91}
Marsden B.~G.,  1991, \aj, 102, 1539

\bibitem[\protect\citeauthoryear{Meeus}{Meeus}{1998}]{meeus98}
Meeus J.,  1998, Astronomical Algorithms.
Willmann-Bell Inc.

\bibitem[\protect\citeauthoryear{Micheli, Bernardi \& Tholen}{Micheli
  et~al.}{2008}]{micheli08}
Micheli M.,  Bernardi F.,    Tholen D.~J.,  2008, \mnras, 390, L6

\bibitem[\protect\citeauthoryear{Needham \& Lu}{Needham \&
  Lu}{1966}]{needham66}
Needham J.,  Lu G.-D.,  1966, Physis, 8, 137

\bibitem[\protect\citeauthoryear{Needham, Lu, Combridge \& Major}{Needham
  et~al.}{1986}]{needham86}
Needham J.,  Lu G.-D.,  Combridge J.~H.,    Major J.~S.,  1986, The Hall of
  Heavenly Records: Korean Astronomical Instruments and Clocks 1380 - 1780.
Cambridge Univ. Press

\bibitem[\protect\citeauthoryear{Park}{Park}{1998}]{park98}
Park C.,  1998, Journal of the Korean History of Science Society, 20, 113

\bibitem[\protect\citeauthoryear{Rufus}{Rufus}{1912}]{rufus12}
Rufus W.~C.,  1912, \tkbras, 4, 23

\bibitem[\protect\citeauthoryear{Rufus}{Rufus}{1936}]{rufus36}
Rufus W.~C.,  1936, \tkbras, 26, 1

\bibitem[\protect\citeauthoryear{Shelton}{Shelton}{1965}]{shelton65}
Shelton J.~W.,  1965, \aj, 70, 337

\bibitem[\protect\citeauthoryear{Stephenson \& Yau}{Stephenson \&
  Yau}{1980}]{stephenson87}
Stephenson F.~R.,  Yau K. K.~C.,  1980, QJRAS, 28, 431

\bibitem[\protect\citeauthoryear{Williams \& Collander-Brown}{Williams \&
  Collander-Brown}{1998}]{williams98}
Williams I.~P.,  Collander-Brown S.~J.,  1998, \mnras, 294, 127

\bibitem[\protect\citeauthoryear{Williams, Ryabova, Baturin \&
  Chernitsov}{Williams et~al.}{2004a}]{williams04b}
Williams I.~P.,  Ryabova G.~O.,  Baturin A.~P.,    Chernitsov A.~M.,  2004a,
  \mnras, 355, 1171

\bibitem[\protect\citeauthoryear{Williams, Ryabova, Baturin \&
  Chernitsov}{Williams et~al.}{2004b}]{williams04a}
Williams I.~P.,  Ryabova G.~O.,  Baturin A.~P.,    Chernitsov A.~M.,  2004b,
  Earth, Moon, and Planets, 95, 11

\bibitem[\protect\citeauthoryear{Williams \& Wu}{Williams \&
  Wu}{1993}]{williams93}
Williams I.~P.,  Wu Z.,  1993, \mnras, 264, 659

\bibitem[\protect\citeauthoryear{Wu \& Williams}{Wu \& Williams}{1992}]{wu92}
Wu Z.,  Williams I.~P.,  1992, \mnras, 259, 617

\bibitem[\protect\citeauthoryear{Yang, Park \& Park}{Yang
  et~al.}{2003}]{yang03}
Yang H.~J.,  Park C.,    Park M.~G.,  2003, Icarus, 175, 215

\end{thebibliography}

\bsp 

\label{lastpage}

\end{document}